\documentclass[twocolumn,showpacs,preprintnumbers,amsmath,amssymb,prl]{revtex4}

\usepackage{graphicx}
\usepackage{dcolumn}
\usepackage{bm}

\begin{document}

\title{Polarization-Correlated Photon Pairs from a Single Quantum Dot}

\author{Charles Santori}
 \email{chars@stanford.edu}
\author{David Fattal}
\author{Matthew Pelton}
\author{Glenn S. Solomon}
 \altaffiliation[Also at ]{Solid-State Photonics Laboratory, Stanford University.}
\author{Yoshihisa Yamamoto}
 \altaffiliation[Also at ]{NTT Basic Research Laboratories, Atsugishi, Kanagawa, Japan.}
\affiliation{
Quantum Entanglement Project, ICORP, JST, E.L.~Ginzton Laboratory, Stanford University,
    Stanford, California 94305
}

\date{\today}

\begin{abstract}
Polarization correlation in a linear basis, but not entanglement, is
observed between the biexciton and single-exciton photons emitted by
a single InAs quantum dot in a two-photon cascade.  The results are
well described quantitatively by a probabilistic model that includes
two decay paths for a biexciton through a non-degenerate pair of
one-exciton states, with the polarization of the emitted photons
depending on the decay path.  The results show that spin
non-degeneracy due to quantum-dot asymmetry is a significant
obstacle to the realization of an entangled-photon generation device.
\end{abstract}


\maketitle

New nonclassical light sources are needed for recently proposed
optical implementations of quantum cryptography~\cite{qcrypt1}
and quantum computation~\cite{qcomp}.
Single semiconductor quantum dots~\cite{generaldot} are attractive as
nonclassical light sources because they have engineered properties,
do not suffer from photobleaching effects, and can be integrated
into larger structures to make monolithic devices.  Quantum dots
have already shown potential as single-photon
sources~\cite{singlephot1,singlephot2,singlephot3}, but they can also generate
sequences of photons in a radiative cascade~\cite{cascade1,cascade2}.
In such a cascade, each photon has a unique wavelength, and the photons
may also have correlated, or even entangled~\cite{benson} polarizations.

In the two-photon cascade, a biexciton singlet state (two electrons
and two holes, 2X) decays to one of two optically active
single-exciton states (one electron and one hole, 1X) by emitting
one photon, and then to the empty-dot state by emitting a second
photon.  In theory, the polarization properties of photon pairs
emitted through these two decay paths result entirely from properties
of the optically-active 1X doublet~\cite{srules1,srules2,srules3}.
For a symmetric quantum dot, the two 1X states are degenerate,
and the two decay paths become ``indistinguishable,'' ideally
producing polarization-entangled photons~\cite{benson}.
For an asymmetric quantum dot, the 1X doublet is split through
the electron-hole exchange interaction into states that couple
to photons having orthogonal linear
polarizations~\cite{xchange,kulakovskii,takagahara}.
If this splitting is much larger than the radiative linewidth,
then the two paths become ``distinguishable,'' with one decay
path producing two horizontally-polarized photons and the other
producing two vertically polarized photons, for example.  In this
case, polarization correlation is expected only in a single,
preferred basis.  Some additional factors are also important,
such as spin flip~\cite{flip} and decoherence
processes~\cite{defaz2,defaz3,defaz4} that randomize the intermediate
1X state, and valence-band mixing~\cite{pseudo}.

In this article, we present an experimental study of the polarization
correlation properties of photon pairs emitted through biexciton
decay in a single InAs quantum dot.  While temporal correlations between
1X and 2X photons have previously been seen~\cite{gerard},
polarization correlation has not yet been reported, to our knowledge,
although a lack of such correlation has been mentioned
elsewhere~\cite{kiraz}.  For our sample, we observe a strong
polarization correlation in a linear polarization basis,
verifying the theoretical picture described above.  However, we do
not observe entanglement, suggesting that quantum-dot asymmetry
is an obstacle to realizing an entangled photon source.

A sample was fabricated containing self-assembled InAs quantum
dots~\cite{generaldot} grown by molecular-beam epitaxy on a (001)
GaAs substrate, capped by $75 \, {\rm nm}$ of GaAs.  A high growth
temperature increased intermixing between the InAs and surrounding
GaAs, shortening the quantum-dot emission wavelengths. 
Mesas about $120 \, {\rm nm}$ tall, $200 \, {\rm nm}$
wide, and spaced $50 \, \mu{\rm m}$ apart were fabricated by electron-beam
lithography and dry etching.  The dots are sparse enough (11 $\mu$m$^{-2}$)
that the smallest mesas contain, on average, fewer than one dot.

\begin{figure}
\includegraphics[width=3.7in]{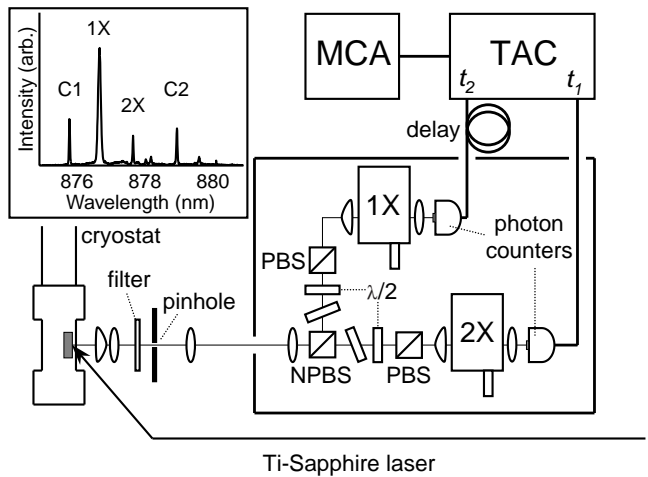}
\caption{\label{fig1} Photon correlation setup: A modelocked Ti-Sapphire
laser excites the sample inside of a cryostat.  The collected emission
is split into two arms by a nonpolarizing beamsplitter (NPBS).  The two
arms count 1X and 2X photons, with rotatable half-wave plates ($\lambda/2$)
followed by polarizers (PBS) determining the measurement polarizations.
An electronic system (delay, TAC, MCA) generates the photon correlation
histogram.
{\it Inset:} single-dot emission spectrum under CW, $650 \, {\rm nm}$
excitation, showing the single-exciton (1X), biexciton (2X), and
charged-exciton (C1, C2) lines.}
\end{figure}

The setup used to characterize the quantum-dot emission is shown in
Fig.~\ref{fig1}.  The sample was cooled to 3-5K in a cryostat.
Single mesas were excited above the GaAs bandgap ($710 \, {\rm nm}$)
by horizontally polarized, $3 \, {\rm ps}$ Ti-Sapphire laser pulses
every $13 \, {\rm ns}$, using a beam incident $54^\circ$ from normal
and focussed to a $20 \, \mu{\rm m}$ spot size on the sample surface.
The emission from the dot was collected by an NA=0.5 aspheric
lens, spectrally filtered to remove laser scatter, imaged onto a
pinhole that selects emission from a $7 \, \mu{\rm m}$-wide region of
the sample, and sent to a Hanbury Brown and Twiss-type (HBT)
correlation setup.  The HBT setup begins with a nonpolarizing beamsplitter,
followed by retarders to correct for the polarization-dependent phase shifts
caused by this beamsplitter.  Each arm then includes a rotatable half-wave plate
to select the measurement polarization, a horizontal polarizer, a small
monochromator ($0.35 \, {\rm nm}$ resolution),
and an avalanche-photodiode photon counter
(EG\&G SPCM) having about $200 \, {\rm s}^{-1}$ dark counts.  
The electrical pulses from the photon counters served as ``start'' ($t_1$) and
``stop'' ($t_2$) triggers for a time-to-amplitude converter (TAC), whose output
was converted to a histogram by a multi-channel analyzer (MCA) card in a computer.  The
resulting histogram of time intervals $\tau = t_2 - t_1$ is equivalent to
a measurement of the photon correlation function, since the collection
efficiency is extremely low.  The ``stop'' signal was delayed by
$100 \, {\rm ns}$ to allow both negative and positive values of $\tau$ in
the correlation histogram.

A photoluminescence spectrum of the dot chosen for this study under
continuous-wave (CW), above-band excitation ($650 \, {\rm nm}$)
is shown in the inset of Fig.~\ref{fig1}.
The lines labeled 1X and 2X are identified as one-exciton
and biexciton emission, respectively, while the lines labeled C1
and C2 are identified as charged-exciton~\cite{chargeX} emission.  
For all of the correlation measurements to be presented, the
``start'' counter was tuned to the 2X line, and the ``stop'' counter was
tuned to the 1X line.

This quantum dot has a large polarization anisotropy.  For convenience,
we designate ``H'' to be a linear polarization rotated $18^\circ$ from
the horizontal lab axis (chosen to maximize the observed polarization
correlations), and ``V'' as the orthogonal polarization.
The horizontal lab axis is aligned with one of the two GaAs
cleave directions, (110) or (1-10).
The photon count
rate for H is nearly double that for V.  The normalized Stokes vectors
of the 2X and 1X lines are $\mathbf{S}_{2X} = (0.34, -0.09, -0.08)$ and
$\mathbf{S}_{1X} = (0.28, -0.12, 0.04)$, respectively, where the three
components are the intensity visibilities in the H/V,
$+45^{\circ}$/$-45^{\circ}$, and circular bases, respectively.  Most dots
on this sample have polarization anisotropy, though the direction of
the Stokes vector varies.  Such anisotropy has been reported
elsewhere~\cite{aniso1,aniso2}, and is related to the asymmetry of the
quantum dot or its environment.

\begin{figure}
\includegraphics[width=3.3in]{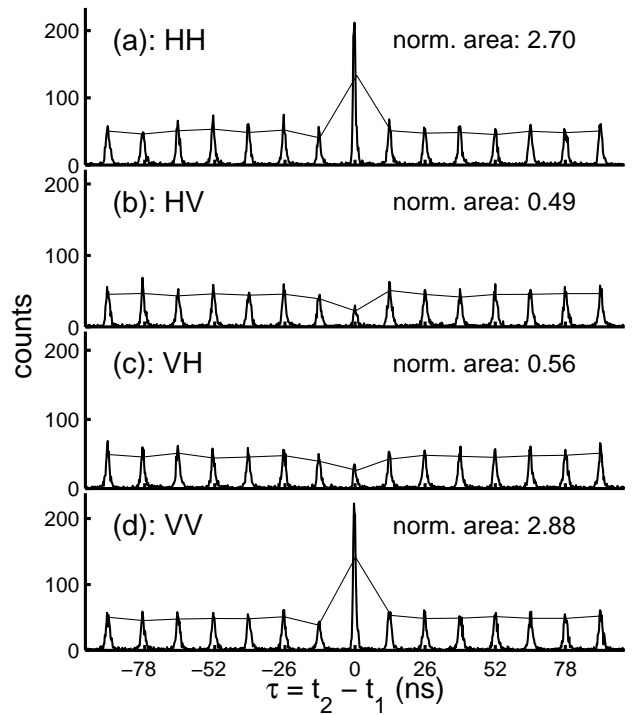}
\caption{\label{fig2} Photon correlation histograms between the
2X and 1X emission lines for four polarization combinations:
(a) HH, (b) HV, (c) VH, and (d) VV, where the first and second letters refer
to the 2X and 1X polarizations, respectively.
H is a linear polarization rotated $18^{\circ}$ from lab horizontal,
and ${\rm V} \perp {\rm H}$.  The central peak at $\tau = 0$ results
from 2X-1X coincidences, and its area, normalized relative to the side
peak average, is indicated.  The solid lines indicate relative peak areas.
The large HH and VV central peak areas and small HV and VH areas
demonstrate polarization correlation.}
\end{figure}
Photon correlation histograms for four special polarization combinations
are shown in Fig.~\ref{fig2}.  The histograms display a series of peaks,
separated by the laser repetition period.  Counts in the central peak at
$\tau = 0$ occur when both a 2X photon and a 1X photon are detected following
the same laser pulse, and its area is proportional to the 2X-1X coincidence rate.
Counts in the side peaks occur when two photons are detected that resulted from
different laser pulses.  The side peaks far from $\tau = 0$
provide an uncorrelated normalization standard, with areas proportional
to the product of the 2X and 1X count rates.  The integration times were
chosen to yield approximately the same side peak area for each histogram.
A rise of the side peaks near $\tau=0$ as reported in \cite{singlephot1}
does not appear here, because the excitation energy here is above the
GaAs bandgap.

\begin{figure}
\includegraphics[width=3.3in]{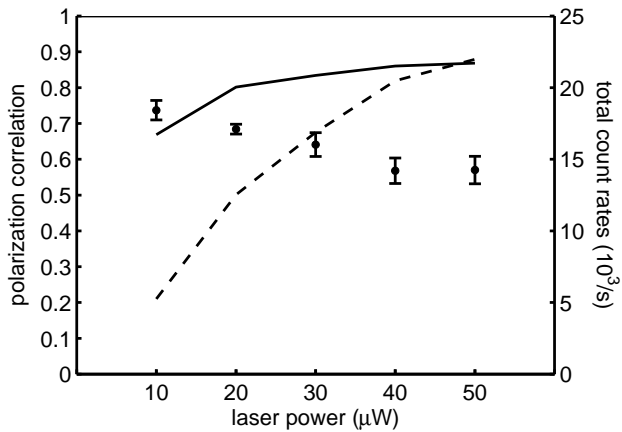}
\caption{\label{fig3} Left axis: measured polarization correlation
function $\chi_{HV}$, as defined in the text (Eq.~\ref{chieq}), versus
pump laser power (points).  The error bars include only Poisson photon
number statistics.  Right axis: total count rates (summed over orthogonal
polarizations) for 1X emission (solid line) and 2X emission (dashed line).}
\end{figure}
It is clear from Fig.~\ref{fig2} that, in the chosen measurement basis, a large
degree of polarization correlation exists between the 2X and 1X photons.
The central peak appears much larger relative to the side peaks for the
${\rm H}_{\rm 2X}$/${\rm H}_{\rm 1X}$ and ${\rm V}_{\rm 2X}$/${\rm V}_{\rm 1X}$
cases (a,d) than for the ${\rm H}_{\rm 2X}$/${\rm V}_{\rm 1X}$ and
${\rm V}_{\rm 2X}$/${\rm H}_{\rm 1X}$ cases (b,c).  We choose to quantify
the degree of correlation by the function
\begin{equation}
\label{chieq}
\chi_{HV} = \frac{\sqrt{C_{\rm HH} C_{\rm VV}} - \sqrt{C_{\rm HV} C_{\rm VH}}}
            {\sqrt{C_{\rm HH} C_{\rm VV}} + \sqrt{C_{\rm HV} C_{\rm VH}}} \, ,
\end{equation}
where $C_{\alpha \beta}$ is the coincidence rate for 2X and 1X measurement
polarizations of $\alpha$ and $\beta$, respectively.
This function yields
values of $+1$, $0$ and $-1$ for the cases of perfect polarization correlation,
independent polarizations, and perfect polarization anticorrelation, respectively,
and is simply related to the polarization-flip probability $\epsilon$
in the model described below.  The data shown in Fig.~\ref{fig2}
were acquired with $20 \, \mu{\rm W}$ excitation power, producing a correlation
of $\chi_{HV} = 0.684$.  The measured value of $\chi_{HV}$ for a range of excitation powers
is plotted in Fig.~\ref{fig3}.  The 2X and 1X count rates are also shown.
A large degree of correlation occurs even when the 2X and 1X count rates
are close to saturation.

While a strong polarization correlation is seen in the H/V basis, negligible
correlation is seen in the $+45^{\circ}$/$-45^{\circ}$ basis
($\chi_{+45,-45} = 0.055$),
suggesting that the photon pairs have negligible entanglement.  To obtain the
entire two-photon density matrix, we followed the procedure outlined
in~\cite{kwiat}.  The density
matrix can be determined from the coincidence rates measured with the
following 2X-1X polarization combinations:
HH, HV, VH, VV, HD, HL, DH, RH, DD, RD, RL, DR, DV, RV, VD, and VL, where the
first and second letters refer to the 2X and 1X measurement polarizations,
respectively, D refers to $+45^{\circ}$, and R and L are the
orthogonal circular polarizations.  For measurement combinations including
a circular polarization, a single quarter-wave plate was inserted after
the collection lens.  To minimize the effect of sample position drift, a
significant error source, we calculated the coincidence rate from the ratio
of the central correlation peak area to the more distant side peak areas: 
\begin{equation}
\label{sidenorm}
D_{\alpha \beta} =  \frac{C_{\alpha \beta}(0)}{\bar{C}_{\alpha \beta}(\tau_{n})}  
                    \, \frac{1 + \mathbf{S}_{2X} \cdot \mathbf{M}_{2X}}{2}
                    \, \frac{1 + \mathbf{S}_{1X} \cdot \mathbf{M}_{1X}}{2} \, ,
\end{equation}
where $D_{\alpha \beta}$ is the corrected coincidence rate for 2X and 1X polarizations
$\alpha$ and $\beta$, respectively, $C_{\alpha \beta}(0)$ is the raw
central peak area, $\bar{C}_{\alpha \beta}(\tau_n)$ is the mean area of
the more distant side peaks, $\mathbf{S}_{2X}$ and $\mathbf{S}_{1X}$ are the 2X and
1X normalized Stokes vectors, and $\mathbf{M}_{2X}$ and $\mathbf{M}_{1X}$ are the
Stokes vectors of the 2X and 1X measurement polarizations.  The effect of sample drift
is largely canceled, since $C_{\alpha \beta}(0)$ and $\bar{C}_{\alpha \beta}(\tau_n)$
both depend on the square of the collection efficiency.  The polarization dependence
of $\bar{C}_{\alpha \beta}(\tau_n)$ is canceled by
the last two terms.

\begin{figure}
\includegraphics[angle=-90,width=3.3in]{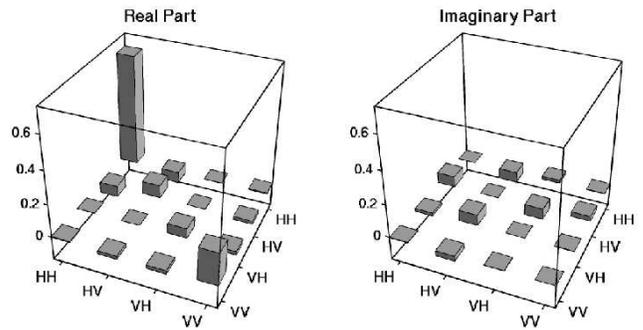}
\caption{\label{fig4} Graphical representation of the two-photon
polarization density matrix describing the 2X and 1X collected photons
under $20 \, \mu{W}$ excitation power.}
\end{figure}
The normalized density matrix obtained for a pump power of $20 \, \mu{\rm W}$
is shown in Fig.~\ref{fig4}.  The relatively small off-diagonal elements
show that little, if any, entanglement is present.  This matrix can
in fact be shown by the Peres criterion to be separable~\cite{peres}.
The on-diagonal components display the polarization correlation that appears
in the H/V basis.  Their values are given in Table~\ref{denstab}.

To model our results, we consider a two-path decay process for the initial biexciton,
as described in~\cite{kulakovskii}.  In one path, both photons are emitted
with $\pi_x$ polarization (detected as H), while in the other path, both are
$\pi_y$ (detected as V).  We assume that the two paths occur
with equal probability.  We also include a probability $\epsilon$ that the
two photons have opposite polarizations.
This takes into account both spin flipping in the one-exciton state, and the
possibility of non-ideal selection rules.  The coincidence probabilities are then
$C_{\alpha \alpha} = \eta_{\alpha,2X} \, \eta_{\alpha,1X} \, (1 - \epsilon) / 2$ and
$C_{\alpha \beta}  = \eta_{\alpha,2X} \, \eta_{\beta,1X} \, \epsilon / 2$, where
$\alpha$ represents either H or V polarization, $\beta \perp \alpha$, and
$\eta_{\alpha,2X}$ and $\eta_{\alpha,1X}$ are the 2X and 1X polarization-dependent
collection efficiencies.  To compare this model to the data, we calculate
the collection efficiency ratios from the measured Stokes vectors, obtaining
$\eta_{V,2X} / \eta_{H,2X} = 0.494$ and $\eta_{V,1X} / \eta_{H,1X} = 0.560$.
To estimate $\epsilon$, we note that $\epsilon = (1 - \chi_{HV}) / 2 = 0.158$, where
$\chi_{HV}$ is defined in Eq.~\ref{chieq}.  The on-diagonal density matrix elements predicted
using these parameters are given in Table~\ref{denstab}, and are in close agreement with
the measured values.  From this model, we can infer that the 1X polarization
flip time $T_1$ is at least 
$\tau_{rad} (1-2\epsilon) / \epsilon = 2.2 \, {\rm ns}$,
where $\tau_{rad} = 0.5 \, {\rm ns}$ is the 1X recombination lifetime. 

The large difference between $\rho_{HH,HH}$ and $\rho_{VV,VV}$ is related
to the unequal detection rates of H and V photons.  In the model above, we assume
that this is due to different collection efficiencies for $\pi_x$ and $\pi_y$
photons.  One might alternatively assume preferential decay through
the $\pi_x$ path, but this explanation cannot simultaneously predict
the measured two-photon density matrix and the measured single-photon
Stokes vectors.  The collection efficiencies for $\pi_x$ and $\pi_y$
must be different, perhaps due to the different angular dipole radiation
patterns for $\pi_x$ and $\pi_y$ photons~\cite{pseudo}.
\begin{table}
\caption{\label{denstab}On-diagonal elements of the two-photon polarization density
matrix, measured and predicted.}
\begin{ruledtabular}
\begin{tabular}{lcccc}
           & $\rho_{HH,HH}$ & $\rho_{HV,HV}$ & $\rho_{VH,VH}$ & $\rho_{VV,VV}$  \\
  Measured & 0.669          & 0.078          & 0.059          & 0.194 \\
  Model    & 0.678          & 0.071          & 0.063          & 0.188 \\
\end{tabular}
\end{ruledtabular}
\end{table}

The fact that we see polarization correlation in only one linear basis and
not entanglement suggests that quantum-dot asymmetry is a dominant effect.
Neglecting spin relaxation, we calculate that the reduced off-diagonal density
matrix element $\rho_{HH,VV}$ produced by an ideal two-photon cascade is
$0.5 / (1 + i \Delta \omega \, \tau_{rad})$, where $\Delta \omega$ is the
frequency splitting of the 1X state, and $\tau_{rad}$ is the 1X radiative
lifetime.  When $\Delta \omega \, \tau_{rad} \gg 1$, $\rho_{HH,VV}$
vanishes, and the entanglement disappears.  This is expected, since the
two decay paths can then be distinguished by the energies of the emitted
photons.  We infer from the fact that we do not see significant entanglement
that $\hbar \Delta \omega \gg 1.3 \, \mu{\rm eV}$, using
$\tau_{rad} = 0.5 \, {\rm ns}$, but we also know
that $\hbar \Delta \omega < 50 \, \mu{\rm eV}$, since we cannot resolve the
polarization splitting spectrally.  Since similar spin splittings have
been reported for other material systems~\cite{defaz4,aniso2},
it appears that spin splitting will be a major obstacle to the
realization of an entangled-photon device.

Several possible remedies might be used to reduce the ratio of the
1X energy splitting to the spontaneous emission rate. 
One could attempt to reduce $\Delta \omega$
by optimizing quantum-dot growth methods, and for this purpose a
systematic study would be useful.  It may also be possible to
force the 1X states of an asymmetric quantum dot into degeneracy
by applying an electric field or a strain to the sample.
Spin relaxation would remain an issue, but a recent experimental
result shows that the 1X spin decoherence time $T_2$ can be much
longer than the radiative lifetime in CdSe quantum
dots~\cite{defaz4}.  Alternatively, one could reduce $\tau_{rad}$, either by
using larger quantum dots~\cite{bellessa}, or by placing the 1X state
on resonance with a microcavity~\cite{gerard2}.  It has already been
demonstrated that the spontaneous emission rates of quantum dots
can be enhanced by a factor of up to 5 in pillar
microcavities~\cite{solomon}.  We hope that by one or more of
these methods, a quantum-dot source of entangled photons
may eventually be demonstrated.

\begin{acknowledgments}
The authors thank H.~Kamada and O.~Benson for valuable discussions.  This
work is partly supported by MURI DAAD19-00-1-0172 (UCLA).  G.S.S. acknowledges
support from DARPA, ARO and JST.
\end{acknowledgments}

\end{document}